# Dynamic conductivity of doped graphene in post-linear response

**B.M. Ruvinskii**[*,1] **and M.A. Ruvinskii**[2]

[1] Department of General and Applied Physics, National Technical University of Oil and Gas, Ivano-Frankivsk, 76000, Ukraine
[2] Department of Physics and Chemistry of Solid State, Precarpathian National University, Ivano-Frankivsk, 76000, Ukraine

* Corresponding author: e-mail bruvinsky@gmail.com

The interband and intraband conductivities of doped graphene were theoretically investigated beyond the linear response. The new dependences of induced currents on frequency and amplitude of external electric field, the graphene temperature and chemical potential were determined for sufficiently strong electric fields in the resonant approximation. Particularly, the saturation of induced currents and the non-linear increase of optical transparency with wave intensity growth were obtained for arbitrary temperatures and doped situation. As contrasted to increase for the interband transitions at fixed intensity, the transmission coefficient of intraband mechanism decreases with rise of the chemical potential and temperature.

**1 Introduction** A number of theoretical works [1-3] revealed universal behavior of low-temperature dynamic conductivity and optical transparency of graphene monolayer as a linear response of Dirac fermions to alternating electric field and incident electromagnetic wave. This behavior is mainly determined by universal physical constants and fine structure constant, it was experimentally confirmed in the infrared [4-6] and visible [7] spectral region. In the letter [8] the nonlinear and resonant response was related only with the particular case of interband transitions in clean graphene at zero absolute temperature and chemical potential $\theta = \mu = 0$.

The present article contains different and more general method of consideration for the interband and inraband conductivity of doped graphene in post-linear and resonant response at arbitrary temperature and chemical potential. We apply the density matrix theory for the quantum transitions in doped graphene. In our work the approximate expression for the function of distribution of stationary nonequilibrium state of Dirac electrons at sufficiently strong electric fields was derived, that it is useful for the other problems of physical kinetics.

The obtained results of present work correspond to the resonant approximation known in quantum optics as the rotating wave approximation.

The new modified nonlinear effects of saturation of both dissipative and non-dissipative parts of induced current were obtained under general conditions, as well as relevant "nonuniversal" features from effects of doping, finite temperature, interband and intraband transitions for the graphene transparency behavior at sufficiently large intensity of incident radiation.

**2 Model** The Hamiltonian of system in the vicinity of Dirac points within pseudo-spin space of graphene sublattices:

$$\hat{H} = u\hat{\boldsymbol{\sigma}}\left(\hat{\boldsymbol{p}} - \frac{e}{c}\boldsymbol{A}\right) = \hat{H}_0 + \hat{V}, \qquad (1)$$

where $\hat{H}_0 = u\hat{\boldsymbol{\sigma}}\hat{\boldsymbol{p}}$, $\hat{\boldsymbol{\sigma}}$ are Pauli matrices, $u$ is Fermi speed, typical for graphene, $\hat{\boldsymbol{p}} = -i\hbar\nabla$, $e$ is electron charge, $\boldsymbol{A}(t)$ is vector potential of uniform alternating electric field.

The electric field (with a frequency $\omega$) $\boldsymbol{E}(t) = \boldsymbol{E}_0\exp(-i\omega t)$ is directed along the x-axis and causes appearance of ac current of density $\boldsymbol{j}$ in the graphene. The operator of current density $\hat{\boldsymbol{j}} = eu\hat{\boldsymbol{\sigma}}$. The eigenvalues $\varepsilon = sup$ and eigenfunctions [9]:

$$|\mathbf{p},s\rangle = \frac{1}{L}\exp\left(\frac{i}{\hbar}\mathbf{p}\mathbf{r}\right)\cdot\frac{1}{\sqrt{2}}\begin{pmatrix} s \\ e^{i\varphi_p} \end{pmatrix}, \quad (2)$$

correspond to operator $\hat{H}_0$; $s=+1$ for 1 state (of the electron within conduction band) and $s=-1$ for 2 state (of the electron within valence band), $p_x = p\cos\varphi_p$, $p_y = p\sin\varphi_p$, $\varphi_p$ is the angle between two-dimensional vector of electron momentum and electric field, $L^2$ is the system area.

In the graphene conductivity the corrections of electron-electron interaction are small [10] except for the excitonic effects. We take into account the finite lifetime $1/\Gamma_p$ of carriers, which is assumed to be the same for electrons and holes due to electron-hole symmetry. For the quasiparticle picture to be meaningful, it is necessary to have $\Gamma \ll \omega$ in any case.

As a result of electric field effect, $\varepsilon$-energy is shifted ($\Delta\varepsilon$) during an electron lifetime in the conduction band. According to the probability multiplication theorem, the corresponding distribution function $\rho_1(\varepsilon,\Delta\varepsilon)$ can be expressed as:

$$\rho_1(\varepsilon,\Delta\varepsilon) = P(\varepsilon,\Delta\varepsilon) f_0(\varepsilon-\mu), \quad (3)$$

where $P(\varepsilon,\Delta\varepsilon)$ is conditional distribution function, $f_0(\varepsilon-\mu) = (\exp[(\varepsilon-\mu)/\theta]+1)^{-1}$ is Fermi-Dirac function, $\mu$ is chemical potential of graphene in equilibrium state. The conditional distribution function can be defined, taking into account the time factor of attenuating states $\exp(-\Gamma_p t)$,

$$P(\Delta\varepsilon\,|\,\varepsilon) = \Gamma_p \,\mathrm{Re}\int_0^\infty \exp\left(\frac{i}{\hbar}\Delta\varepsilon t - \Gamma_p t\right)dt = \frac{(\hbar\Gamma_p)^2}{(\Delta\varepsilon)^2 + (\hbar\Gamma_p)^2} \quad (4)$$

and $(\Delta\varepsilon)^2$ can be found from the amplitude of diagonal matrix element $V_{11}$ of intraband transition:

$$(\Delta\varepsilon)^2 = |V_{11}|^2 = \left(\frac{eup_x}{\omega p}E_0\right)^2. \quad (5)$$

From (3)-(5) we obtain a formula for the function of distribution of stationary non-equilibrium state of an electron within conduction band $\rho_1(\varepsilon,\Delta\varepsilon) \equiv \rho_1^0$:

$$\rho_1^0 = \left[1+\zeta^2\left(\frac{p_x}{p}\right)^2\right]^{-1} \cdot f_0(\varepsilon-\mu), \quad (6)$$

where

$$\zeta = \frac{euE_0}{\hbar\omega\Gamma_p} \quad (7)$$

is the typical dimensionless parameter of the task. Similarly we have for the distribution function within valence band:

$$\rho_2^0 = \left[1+\zeta^2\left(\frac{p_x}{p}\right)^2\right]^{-1} \cdot f_0(-\varepsilon+\mu), \quad (8)$$

where $f_0(-\varepsilon+\mu) = (\exp[(-\varepsilon+\mu)/\theta]+1)^{-1}$. Apart from $\Gamma \ll \omega$, (5) also results in limiting the upper value $\zeta$ (or electric field) $\zeta \ll \omega/\Gamma$.

We have derived the following equation for chemical potential of graphene in equilibrium state:

$$n_0 = \frac{2}{\pi(\hbar u)^2}\left\{\frac{\mu^2}{2} + \theta^2\left[\frac{\pi^2}{6} + 2\sum_{m=1}^\infty \frac{(-1)^m}{m^2}\exp\left(-\frac{m\mu}{\theta}\right)\right]\right\} \quad (9)$$

($n_0$ is concentration of electrons and holes), where $\mu \approx \hbar u\sqrt{\pi n_0}$ at sufficiently low temperature $\theta \ll \mu$, and $\mu \approx \pi(\hbar u)^2 n_0 / 4\theta\ln 2$ at high temperature $\theta \gg \mu$, as consistent with [3]. The Eq.(9) coincides exactly with the Eq.(74) of the paper [11] in case of zero gap $\Delta = 0$ by using identities for dilogarith function.

**3 The equation for the density matrix and quantum transitions in doped graphene** The quantum equation of motion for the statistical operator $\hat{\rho}$ (density matrix)

$$i\hbar\frac{\partial\hat{\rho}}{\partial t} = \hat{H}\hat{\rho} - \hat{\rho}\hat{H} \quad (10)$$

is written for the matrix elements of the interband transitions ($\rho_{12}$ for the $2\to 1$ transition)

$$i\hbar\frac{\partial\rho_{12}}{\partial t} = (H_{11}-H_{22})\rho_{12} + (\rho_{22}-\rho_{11})H_{12}, \quad (11)$$

where "1" is index of the electron state in the conduction band, and the "2" in the valence band. For the basic system functions we choose the eigenfunctions (2) of operator $\hat{H}_0$ in the absence of an external electric field. Interband transition is a vertical transition in which the two-dimensional vector $\mathbf{p}$ is an integral of motion. The change in the unperturbed energy of the electron in the interband transition $2\to 1$ is equal

$$(H_0)_{11} - (H_0)_{22} = 2\varepsilon, \quad (12)$$

and $\hbar\omega \approx 2\varepsilon$ in conditions close to the exact resonance. The operator $\hat{V}(t)$ for interaction of the electron with the electric field, where the symbol Re is omitted before the complex factor, has form:

$$\hat{V}(t) = i\frac{eu}{\omega}\hat{\sigma}_x E_0 \exp(-i\omega t). \quad (13)$$

Matrix elements of the operator (13) in the base system functions are equal:

$$V_{\alpha\beta}(t) = H_{\alpha\beta} = (-1)^\alpha \frac{eu}{\omega}\cdot\frac{p_y}{p}E_0 e^{-i\omega t} \quad (14)$$

for $\alpha \neq \beta = 1,2$, and

$$V_{\alpha\alpha}(t) = -i \cdot (-1)^\alpha \frac{eu}{\omega} \cdot \frac{p_x}{p} E_0 e^{-i\omega t} \quad (15)$$

for $\alpha = \beta = 1,2$. The matrix elements $\rho_{11}$ and $\rho_{22}$ correspond to the intraband transitions. Then we find the stationary solutions of equations (11) for the interband transitions ($2 \to 1$ and $1 \to 2$) in the resonant interaction, when

$$|\hbar\omega - 2\varepsilon| \ll \hbar\omega, 2\varepsilon \quad (16)$$

and in the stationary regime $2\varepsilon \gg |V_{11} - V_{22}|$ for $t \to \infty$ we take into account only the linear time factor $e^{-i\omega t}$, omitting the terms, oscillating at twice the frequency:

$$\rho_{\alpha\beta}(t) \approx \frac{\rho_{\beta\beta} - \rho_{\alpha\alpha}}{\hbar\omega + 2\varepsilon(-1)^\alpha} V_{\alpha\beta}(t) \quad (\alpha \neq \beta). \quad (17)$$

The diagonal elements $\rho_{\beta\beta}$ and $\rho_{\alpha\alpha}$ ($\alpha \neq \beta$) in formula (17) are not dependent on time and have a sense of intraband distribution functions of the stationary non-equilibrium state. Thus the post-linear response is related to the determination of these functions, depending on the amplitude of the electric field. The explicit form of the diagonal elements is unknown and it can not be obtained only from the equation (11) without the additional information about the subsystem, for example, from the master equation. The system of quantum kinetic equations can be formulated in the $\mu$-space [12] with a single relaxation time $\Gamma^{-1}$ for graphene:

$$i\hbar \frac{\partial \rho_{\alpha\alpha}}{\partial t} = (V_{\alpha\beta}\rho_{\beta\alpha} - \rho_{\alpha\beta}V_{\beta\alpha}) + i\hbar(w_{\alpha\beta}\rho_{\beta\beta} - w_{\beta\alpha}\rho_{\alpha\alpha}), \quad (18)$$

$$i\hbar \frac{\partial \rho_{\alpha\beta}}{\partial t} = [2\varepsilon(-1)^\beta + V_{\alpha\alpha} - V_{\beta\beta} - i\hbar\Gamma]\rho_{\alpha\beta} + (\rho_{\beta\beta} - \rho_{\alpha\alpha})V_{\alpha\beta}, \quad (19)$$

where $\alpha \neq \beta = 1,2$, $w_{\alpha\beta} = \Gamma f_{0\beta}$ is the probability of transition states $\beta \to \alpha$ in a unit of time, $f_{0\beta}$ is the equilibrium distribution function of the charge carriers. Using the well-known methods for finding stationary solutions in the resonant approximation [12,13], from the equation (18) with taking into account (19) at $(\hbar\omega - 2\varepsilon)^2 \ll (\hbar\Gamma)^2$ we get

$$\rho_{\beta\beta} - \rho_{\alpha\alpha} \approx \frac{(\hbar\Gamma)^2}{(\Delta\varepsilon)^2 + (\hbar\Gamma)^2}(f_{0\beta} - f_{0\alpha}), \quad (20)$$

that is consistent with (3)-(8).

**4 Current densities** We use the above simple approximations for the calculation of current densities.

Matrix elements of the interband current density in the eigenfunctions (2) are equal

$$(j_x)_{\alpha\beta} = (-1)^\alpha i \frac{eup_y}{p} \cdot \frac{1}{L^2} \quad (\alpha \neq \beta = 1,2). \quad (21)$$

Matrices (14), (15), (17) and (21) are diagonal in the momentum representation, so the definition of statistical average of the interband current density is determined by taking into account the spin and valley degeneracy $g_s$ and $g_v$ ($g_s = g_v = 2$):

$$\langle j_x \rangle = Sp\hat{\rho}\hat{j}_x = \sum_p \sum_{\alpha,\beta,\alpha\neq\beta} \rho_{\alpha\beta}(j_x)_{\beta\alpha}, \quad (22)$$

where

$$\rho_{12}j_{21} + \rho_{21}j_{12} = i\left(\frac{eup_y}{Lp}\right)^2 (\rho_2^0 - \rho_1^0)E_0 e^{-i\omega t} \times \\ \times \frac{4\varepsilon}{\omega[(\hbar\omega)^2 - (2\varepsilon)^2]}. \quad (23)$$

Note that for finding the finite current $\langle j_x \rangle$ we must subtract in (23) from the expression $4\varepsilon\{\omega[(\hbar\omega)^2 - (2\varepsilon)^2]\}^{-1}$ the same expression at $\omega \to 0$. This known circumstance [2,3] is due to the fact that the current must vanish if vector-potential $A$ is not dependent on the time and coordinates. After this in the obtained expression we have to make the substitution $\omega \to \omega + i\eta$ ($\eta \to +0$), obliged to the adiabatic condition [11]: at $t \to -\infty$ the density matrix reduces to the equilibrium density matrix. As a result, we have

$$\langle j_x \rangle = g_s g_v \sum_p \frac{(eu\hbar)^2}{\varepsilon L^2}\left(\frac{p_y}{p}\right)^2 \frac{\rho_2^0 - \rho_1^0}{\hbar^2(\omega + i\eta)^2 - (2\varepsilon)^2} E_0 e^{-i\omega t}. \quad (24)$$

The formula (24) looks only linear response, but the difference $\rho_2^0 - \rho_1^0$ does not apply the case of equilibrium and it characterizes the non-equilibrium stationary states, depending on amplitude of the external electric field. Further calculations in (24) associated with the replacement $\sum_p \ldots \to (L/2\pi\hbar)^2 \int \ldots dp_x dp_y$ and the use of formula

$$[\hbar^2(\omega + i\eta)^2 - (2\varepsilon)^2]^{-1} = P[(\hbar\omega)^2 - (2\varepsilon)^2]^{-1} - \\ -i\pi\delta[(\hbar\omega)^2 - (2\varepsilon)^2], \quad (25)$$

where $P$ is the symbol of the principal value of the integral.

Taking into consideration (24), (25) and (6)-(8) for dissipative part of interband current, we will obtain:

$$j_d^{\text{inter}}(t) = \frac{2\sigma_0 G(\omega,\theta,\mu)}{\sqrt{1+\zeta^2}+1} E_0 \cos\omega t, \quad (26)$$

$\sigma_0 = e^2/4\hbar$ is the conductance quantum,

$$G(\omega,\theta,\mu) = \text{sh}\left(\frac{\hbar\omega}{2\theta}\right) \cdot \left[\text{ch}\left(\frac{\mu}{\theta}\right) + \text{ch}\left(\frac{\hbar\omega}{2\theta}\right)\right]^{-1}, \quad (27)$$

the change of $\zeta$ shall be made in formula (7) $\Gamma_p \to \Gamma \equiv \Gamma_{\hbar\omega/2u}$, where the momentum $p$ corresponds to frequency $\omega$.

The non-dissipative part of interband current:

$$j_{nd}^{inter}(t) = -\frac{\sigma_0}{\pi(\sqrt{1+\zeta^2}+1)}\left[\ln\frac{(\hbar\omega+2\mu)^2}{(\hbar\omega-2\mu)^2}\right]E_0\sin\omega t \ . \quad (28)$$

The logarithmic singularity at $\hbar\omega = 2\mu$ is cut by temperature (or currier relaxation) with the replacement $(\hbar\omega-2\mu)^2 \to (\hbar\omega-2\mu)^2 + (2\theta)^2$ in (28).

The intraband current density can be determined also from the Boltzmann kinetic equation for relaxation time approach $\tau \equiv 1/\Gamma$

$$j_x^{intra}(t) = -g_s g_v \frac{e^2 E_0 e^{-i\omega t}}{(2\pi\hbar)^2(\Gamma - i\omega)}\int v_x \frac{\partial}{\partial p_x}(\rho_1^0 - \rho_2^0)d^2\mathbf{p}, \quad (29)$$

where $v_x = up_x/p$, $\rho_1^0$ and $\rho_2^0$ are defined by formulas (6)-(8).

The dissipative part of intraband current, which determines Joule heat generation and graphene transparency reduction,

$$j_d^{intra}(t) = \frac{16\sigma_0 g(\omega,\theta,\mu)}{(\sqrt{1+\zeta^2}+1)\pi}E_0\cos\omega t, \quad (30)$$

where

$$g(\omega,\theta,\mu) = \frac{\theta\cdot(\hbar\Gamma)}{(\hbar\omega)^2+(\hbar\Gamma)^2}\cdot\ln\left[2\text{ch}\left(\frac{\mu}{2\theta}\right)\right]. \quad (31)$$

Non-dissipative part

$$j_{nd}^{intra}(t) = \frac{16\sigma_0\omega g(\omega,\theta,\mu)}{(\sqrt{1+\zeta^2}+1)\pi\Gamma}E_0\sin\omega t \ . \quad (32)$$

If $\zeta \ll 1$ (in weak electric fields or at high frequency and short relaxation time $1/\Gamma$), the obtained formulas (26)-(31) conform to the results [1-3] of the linear response theory (first order of the perturbation theory). If $\zeta \gg 1$, i.e. in case of strong electric fields (or low frequency), there occur saturation of amplitudes of all induced currents (both $j_d$ and $j_{nd}$, interband and intraband ones). As concerns the saturation amplitude of dissipative part of interband current in doped graphene, we have as follows:

$$j_{d,\max}^{inter} = \frac{|e|\omega\Gamma}{2u}\cdot G(\omega,\theta,\mu) \ . \quad (33)$$

Formulas (26) and (33) differ from the ones (4)-(5) in Ref. [8] (at $\theta = 0$ and $\mu = 0$) by factor $G(\omega,\theta,\mu)$ (see (27)), where $G(\omega,0,0) = 1$. If $\Gamma \ll \omega$, the ratio between saturation amplitudes of dissipative parts of intraband current and the interband one is

$$\frac{j_{d,\max}^{intra}}{j_{d,\max}^{inter}} = \frac{8}{\pi}\cdot\frac{\theta\cdot(\hbar\Gamma)}{(\hbar\omega)^2}\cdot\frac{\ln\left[2\text{ch}\left(\frac{\mu}{2\theta}\right)\right]}{G(\omega,\theta,\mu)}. \quad (34)$$

In other words, it is essentially dependent of temperature, field frequency, chemical potential of graphene and scattering mechanism of free charge carriers, which determines function $\Gamma(\omega)$ in particular.

**5 Optical transparency of doped graphene**

When applying the above-obtained formulas (26,27) and (30,31) to the dissipative currents $j_d^{inter}(t)$ i $j_d^{intra}(t)$, we can define the optical transparency of suspended graphene in vacuum (or air) according to the scheme of paper [8], where the interband case was researched for $\theta = 0$ and $\mu = 0$. The intraband conductivity is dominant at low frequency of alternating electric field $\omega < \theta/\hbar$ (or sufficiently high temperatures), but the interband conductivity dominates if $\omega > \theta/\hbar$. Let us consider these two cases separately in order to get corresponding transmission coefficients $T^{inter}$ and $T^{intra}$ for the incident radiation, propagating normal to the plane graphene with the linear polarization. We obtain the following algebraic equations for the transmission coefficients $T = E_T^2/E_0^2$ ($E_T$ and $E_0$ are electric fields of transmitted and incident wave):

$$1 + \frac{8\alpha g(\omega,\theta,\mu)}{1+\sqrt{1+T^{intra}\cdot I}} = \frac{1}{\sqrt{T^{intra}}}, \quad (35)$$

$$1 + \frac{\pi\alpha G(\omega,\theta,\mu)}{1+\sqrt{1+T^{inter}\cdot I}} = \frac{1}{\sqrt{T^{inter}}}, \quad (36)$$

where $\alpha = e^2/\hbar c$ is the fine structure constant, $I = (euE_0/\hbar\omega\Gamma)^2$ is the effective dimensionless intensity of incident wave. The "nonuniversal" features of doped graphene are shown in $g(\omega,\theta,\mu)$ and $G(\omega,\theta,\mu)$ (Eqs.(31), (27)) for the considered transitions. Figures 1, 2 demonstrate the dependences of transmission coefficients on the magnitude of incident wave at different values of chemical potential (with relevant concentration of free charge carriers) and temperature, calculated from (35), (36).

As in the case of the interband absorption relating to intrinsic graphene [8], we have nonlinear increase $T$ with rise of effective intensity $I$ of incident wave: first, at low values of $I$ – linearly [7], then – more slowly $T \to 1$, where

$$T^{intra} \approx 1 - \frac{16\alpha}{\sqrt{I}}g(\omega,\theta,\mu) \ , \quad T^{inter} \approx 1 - \frac{2\pi\alpha}{\sqrt{I}}G(\omega,\theta,\mu) \ .$$

If $I$ is fixed, $T^{intra}$ decreases with growth of $\mu$ and $\theta$, as contrasted to growth of $T^{inter}$ (Figures 1, 2). It is explained by corresponding behavior of $g(\omega,\theta,\mu)$ (31), as compared with $G(\omega,\theta,\mu)$ (27).

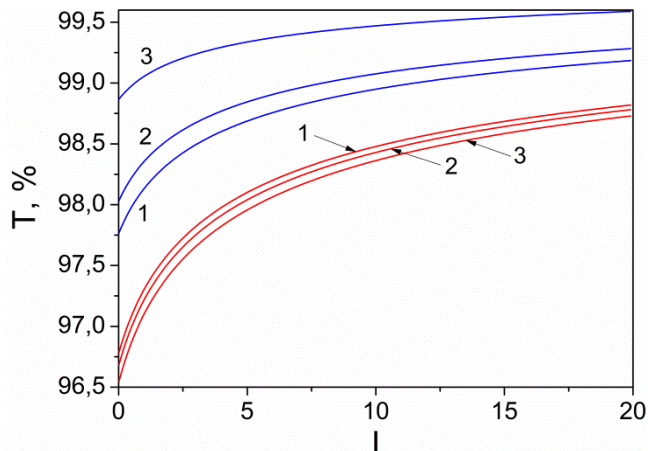 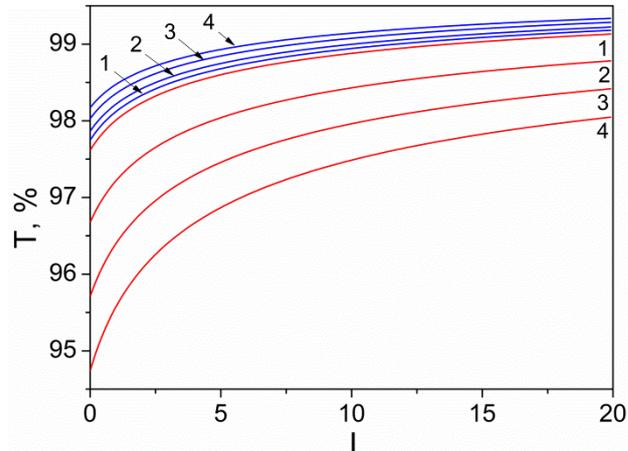

**Figure 1** Dependences of the transmission coefficients $T^{\mathrm{inter}}$ (blue) and $T^{\mathrm{intra}}$ (red) of suspended graphene on the effective intensity of the incident radiation $I = (euE_0/\hbar\omega\Gamma)^2$ at the temperature 300 K. $\hbar\Gamma = 0.010$ eV. The blue curves ($\hbar\omega = 0.3$ eV): 1 – $\mu = 0.01$ eV ($n_0 = 5.3 \times 10^{10}$ cm$^{-2}$); 2 – $\mu = 0.10$ eV ($n_0 = 8.9 \times 10^{11}$ cm$^{-2}$); 3 – $\mu = 0.15$ eV ($n_0 = 1.8 \times 10^{12}$ cm$^{-2}$). The red curves ($\hbar\omega = 0.015$ eV): 1 – $\mu = 0.010$ eV ($n_0 = 5.3 \times 10^{10}$ cm$^{-2}$); 2 – $\mu = 0.015$ eV ($n_0 = 8.1 \times 10^{10}$ cm$^{-2}$); 3 – $\mu = 0.020$ eV ($n_0 = 1.1 \times 10^{11}$ cm$^{-2}$).

**Figure 2** Dependences of $T^{\mathrm{inter}}(I)$ (blue) and $T^{\mathrm{intra}}(I)$ (red) at the fixed $\mu$, $\hbar\omega$ and $\hbar\Gamma = 0.010$ eV for the different temperatures and concentrations $n_0$. The blue curves ($\mu = 0.10$ eV, $\hbar\omega = 0.3$ eV): 1 – 100 K, $n_0 = 7.5 \times 10^{11}$ cm$^{-2}$; 2 – 200 K, $n_0 = 8.1 \times 10^{11}$ cm$^{-2}$; 3 – 300 K, $n_0 = 8.9 \times 10^{11}$ cm$^{-2}$; 4 – 400 K, $n_0 = 1.0 \times 10^{12}$ cm$^{-2}$. The red curves ($\mu = 0.015$ eV, $\hbar\omega = 0.015$ eV): 1 – 200 K, $n_0 = 5.5 \times 10^{10}$ cm$^{-2}$; 2 – 300 K, $n_0 = 8.1 \times 10^{10}$ cm$^{-2}$; 3 – 400 K, $n_0 = 1.1 \times 10^{11}$ cm$^{-2}$; 4 – 500 K, $n_0 = 1.3 \times 10^{11}$ cm$^{-2}$.

**6 Conclusion** In the resonant interaction the obtained dependences of interband and intraband currents of doped graphene on electric field, frequency, chem. potential (of carrier concentration) and temperature prove significant deviations from the linear response theory in sufficiently strong electric fields and at low frequencies. The saturation was defined for all amplitudes of induced currents, as well as for nonlinear increase of appropriate transparency of suspended graphene with intensity growth of incident radiation. The intraband transmission coefficient decreases with chem. potential and temperature rise, while the interband transmission coefficient increases under the same conditions.

The systematic and more detailed experimental research of nonlinear effects of ac conductivity and transparency is topical at high intensity and lower frequencies of incident radiation in the doped and undoped graphene.

**Acknowledgments**

We thank Prof. V.P. Gusynin for important comments and the useful papers.